# Computed rotational spectrum of $C_{70}^{+}$ for detection by radio astronomy

László Nemes *

*Research Center for Natural Sciences, Institute for Materials and Environmental Chemistry, Eötvös Lóránd Research Network* 1519 Budapest, P.O. Box 286, Hungary

**ABSTRACT**

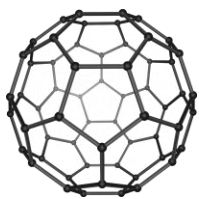

The apolar fullerenes $C_{60}$ and $C_{70}$ are not accessible for radio astronomy. Upon ionization static Jahn-Teller effects occur in $C_{70}^{+}$ that distort the $D_{5h}$ neutral symmetry to $C_s$. This point group is polar thus ionization induces a permanent electric dipole moment in $C_{70}$. The goal of the present calculations is to compute the equilibrium geometry and dipole moment of the $C_{70}^{+}$ cation by various DFT methods and to simulate microwave spectra. Using quantum chemistry rotational constants, Cartesian dipole moment components and the resultant dipole, as well as Jahn-Teller stabilization energies and HOMO-LUMO gaps were obtained. Microwave rotational spectrum simulations for the slightly asymmetric top ion were carried out for gas phase temperatures 2.73 K and 10 K. These spectra may serve as starting point for laboratory microwave measurements and as screening guide in radio astronomical searches. In addition it was found that the static Jahn-Teller effect in $C_{70}^{+}$ is the consequence of the mixing of the two highest ground state occupied orbitals, thus it is a pseudo Jahn-Teller effect.

**Key words:** miscellaneous, molecular data, ISM: molecules, radio lines: ISM

## 1 Introduction

Molecular radio astronomy is mostly pursued for diatomic and comparatively small molecules in mm- and submm ranges. The basic requirement is that the molecule should have a sufficiently large permanent dipole moment. PAH and fullerene molecules cannot usually be detected in these frequency ranges due to their small rotational constants, but mainly because of their apolar molecular structure. Their infrared spectra have however been observed by infrared space telescopes. Following the 1985 laboratory discovery of $C_{60}$ and $C_{70}$ fullerenes (Kroto et al. 1985) the interesting question emerged whether these molecules could be observed in the interstellar space and in various other astronomical objects. This goal was achieved in 2010 (Cami et al. 2010) using the Spitzer infrared telescope observations for the Tc 1 planetary nebula.

*e-mail: nemesl@comunique.hu

Further infrared detections of $C_{60}/C_{70}$ were reported in various other cosmic sources , for example in hydrogen containing planetary nebulae (Garcia-Hernandéz et al. 2010), and in the Magellanic Cloud planetary nebulae (Garcia-Hernandéz et al. 2011). Infrared spectroscopy is an excellent method for detecting large molecules, but when there are several different IR absorbers present, individual molecular identification can be difficult. Theoretical calculations by quantum chemical methods greatly help in this task, but as it has been pointed out recently (Candian et al. 2019) for ionized forms of large, highly symmetric fullerenes, the usual Born-Oppenheimer adiabatic method involved in standard quantum chemistry programs (e.g. Gaussian and Turbomole) , renders it impossible to predict vibrational spectra correctly for the $C_{70}^+$ cation due to vibronic interactions arising from the presence of low lying excited electronic states. This was discussed in a previous quantum chemical and IR experimental work (Kern et al. 2016) . So it would be important to detect these large carbon ions also by radio astronomy.

In addition to Jahn-Teller distortions there are several intramolecular effects that may lead to the emergence of a small electric dipole. Such are vibrational effects in vibrationally excited states, asymmetrically isotope substituted molecules (also a vibrational effect), and centrifugal distortion in highly rotationally excited molecules (that also appears in vibrational ground states). These effects have been experimentally studied in a number of small molecules. For example the dipole moment of mono-deutero methane ($CH_3D$) is about $2.10^{-32}$ C.m, i.e. $6.10^{-3}$ Debye (Ozier et al. 1969). Vibrational effects may induce similarly small dipole moments when two-fold or three-fold degenerate vibrational states are excited, A general treatment of rotational transitions induced by centrifugal distortions was given (Watson 1971) .

There has been one recent work in which it was shown that the singly $^{13}C$ substituted $C_{60}^+$ cation has a dipole moment of about 0.02 Debye (Yamada et al. 2017). The authors point out that the smallness of the predicted dipole moment and the large rotational partition function would make radio astronomical detection very difficult. It should be pointed out that the fully $^{12}C_{60}^+$ ion is completely apolar, as in the next section will be discussed.

## 2. Jahn-Teller geometry distortion in the $C_{70}^+$ cation.

When a molecule is in a degenerate electronic ground state, this state is unstable in the presence of degenerate vibrational states, and the molecule is instantaneously distorted to a lower symmetry, the degeneracy is removed and the molecule is found in a stable lower energy state. Exceptions are linear molecules and Kramers degeneracy. Jahn and Teller published detailed group theoretical results for many symmetry groups (Jahn & Teller 1937). The fundamental requirement for the symmetry of the degenerate vibrational state bringing about the distortion is given by group theory as:

$$[\Gamma(elect)]^2 \cdot \Gamma(s) \supset \Gamma(vib) \qquad (1)$$

where $\Gamma(elect)$ is the symmetry of the degenerate electronic state, $[\Gamma(elect)]^2$ is the symmetric direct product of the representation of the degenerate electronic state, $\Gamma(s)$ is the fully symmetric representation, and $\Gamma(vib)$ is the representation of the Jahn-Teller (JT) active vibrational mode. When the symmetry group of the molecule contains the inversion operation - this is the case for $C_{60}$ that belongs to the icosahedral $I_h$ group - the JT effect in the ion leads to a lower symmetry group ($D_{5d}$) that preserves inversion symmetry thus the ion will not become polar. In spite of the JT effect, no dipole moment arises. Isotopic substitution then can lead to a small dipole as discussed in (Yamada et al. 2017).

In the present case of $C_{70}$ the apolar geometry of the neutral molecule ($D_{5h}$) is distorted upon ionization to $C_s$ symmetry, which is a polar group, thus a permanent dipole is created, as will be shown in the computational section..

A theoretical treatment was given (Child & Longuet-Higgins, 1961) for JT effects in the vibrational and rotational spectra of symmetric top molecules in degenerate electronic ground states. Since $C_{70}$ is a prolate symmetric top and since the $D_{5h}$ symmetry group has two-fold degenerate (e) vibrational states, the Child-Longuet-Higgins treatment predicts a JT-induced electric dipole in the ion. They also showed that when $[\Gamma(elect)]^2$ in Eq.(1) includes $\Gamma_z$ the representation to which the z (figure axis) translation $M_z$ belongs, the induced dipole will be parallel to the z axis, whereas it will be perpendicular to the figure axis ($\mu_{x,y}$) if $[\Gamma(elect)]^2$ includes the doubly degenerate representation to which the (x,y) translations belong. The symmetry of the electronic ground state arising from the JT-effect in the cation can then be obtained from the symmetry of the induced dipole.

First ionization occurs from the frontier (valence) orbital that is also called HOMO (highest occupied molecular orbital) The symmetry of the degenerate electronic ground state that is created upon ionization is usually taken to be the symmetry of HOMO of the neutral species. (e.g. Bendale et al 1992, Ramanantoanina et al. 2005)

However it follows from the Child-Longuet-Higgins symmetry rule (Child & Longuet-Higgins, 1961) that since the JT-induced dipole vector in the present calculations is obtained perpendicular to the figure axis (z) of $C_{70}$ and lies in the (x,y) plane, (see Section 3.), the species of the JT active degenerate vibrational state has symmetry $e'_1$ in the $D_{5h}$ group.. On the other hand DFT calculations for neutral $C_{70}$ yield the HOMO symmetry as $E_1''$ or $a_2''$ in different basis sets. The corresponding JT active vibrations ($e_2'$) are not infrared active but Raman modes. The reason for this discrepacy is explained in Section 4.

## 3. Computational details

Quantum chemical calculations were carried out using Gaussian16 rev.C.01 Linux version (Frisch et al. 2019) and Gauss View 6.1 (Dennington et al. 2019). Linux version as graphical interface. Gaussian 16 was run on Hungarian HPC's (HP SL 250S at the University of Debrecen, HP CP4000 BL at the University of Szeged and SGI UV 2000 at the University of Miskolc). High level post-SCF methods (e.g. QCISD or CCSD) may prove to be off-limits even for other super computers available here, so the calculations were made using open shell unrestricted density functional DFT models. Since the present calculations are for the electronic ground state, and as the Jahn-Teller effect in $C_{70}^+$ is a static one, the adiabatic DFT methods used here should be applicable. In the present calculations the electronic spin–orbit coupling effects were neglected. At any rate if there were strong spin-orbit coupling effects, those would split orbital degeneracy thus remove the Jahn-Teller distortion ( Herzberg 1966).

The starting geometry (Cartesian coordinates) in the optimizations was taken from crystallographic data and geometry refinements were done in Cartesian coordinates. There were no symmetry restrictions applied in the runs and harmonic vibrational analyses were carried out for the ion to detect imaginary frequencies. In the harmonic vibrational analyses no imaginary frequencies were found. The ground state wave functions for the DFT runs in Table 1. were found stable, all optimized geometries corresponded to stationary points.

### 3.1 Discussion of the quantum chemical calculations.

The $C_{70}$ molecule has 204 normal vibrations, and are distributed among the $D_{5h}$ group irreducible representations as follows:

$\Gamma_{vib} = $ 12$A_1'$ + 9$A_1''$ + 9$A_2'$ +10$A_2''$

$+ 21E_1' + 19E_1'' + 22E_2' + 20E_2''$    (2)

Using the symmetry rule Eq.1 it is seen that the (E x e) interactions may involve 19 – 22 vibrational states depending on the HOMO symmetry, Thus the JT interactions in $C_{70}^+$ are formally multimode ones so expected to lead to a rather involved case, even if not all interactions have significant coefficients in the JT

Hamiltonian. The multidimensional potential surface is likely to have many minima in addition to five-fold repetition due to the $D_{5h}$ symmetry of the potential function. As no Jahn-Teller theoretical studies have so far been published for $C_{70}^+$ this problem is still open.

A number of different density functionals and basis sets were used in the present calculations. Those in Table 1. resulted in rather similar results. In order to chose among them for spectral predictions the lowest final SCF energy or the greatest JT stabilisation energy value was considered. From these considerations two methods were taken from Table 1.: the b3lyp/ccpvdz and the pw6b95/ccpvdz methods.

The b3lyp/ccpvdz level provided the largest JT stabilization energy (2069.3 cm$^{-1}$) while the pw6b95/ccpvdz level yielded the lowest final SCF energy : -2671.037109 hartree. To calculate the JT stabilization energy the method in (Bakó et al. 2003) was applied. The difference of two SCF energy values were used, one from optimizing the neutral structure first, and the other obtained from a single point energy calculation for the ion starting from the previously optimized neutral geometry, using the same functional and basis set in both cases. To calculate the JT stabilization energy a correction for the ionization energy is required. In this work the average of the values given in (NIST Chemistry Web Book 2022) : 7.57 eV was used. The HOMO-LUMO gaps in Table 1.are ranging from 20000 to about 23000 cm$^{-1}$. This suggests that the $C_{70}^+$ cation is thermodynamically stable.

The Jahn-Teller interaction in $C_{70}^+$ results in small geometry changes, while the molecule is changed from a prolate symmetric top to a slightly asymmetric top. The reason for small changes is the delocalization of the bonding HOMO orbital over the carbon atom network (see Figure 1.). The rotational constants in Table 1. differ from those in the neutral $C_{70}$ by about 0.1-0.3 % only with the axial constant increasing and the perpendicular two constants decreasing for the b3lyp/ccpvdz model. Similar changes occur for the other DFT/basis set models. For all the models in Table 1. the final geometries are of $C_s$ symmetry.

Table 1. Results of the quantum chemical calculations for $C_{70}^+$

| Method | Point group | Asymmetry parameter($\kappa$) | Dipole moment and components (Debye) | Jahn-Teller Stabilization Energy ($cm^{-1}$) | Rotational Constants (MHz) | $\alpha$HOMO-$\alpha$LUMO gap ($cm^{-1}$) | SCF final converged energy (hartree) |
|---|---|---|---|---|---|---|---|
| b3lyp/6-31+g | $C_s$ | -0.98276 | $\mu$:1.31613 $\mu_x$:-0.77361 $\mu_y$:-1.06477 | 170.90 | A:66.90853 B: 58.09461 C:58.01762 | 19625 | -2666.388998 |
| b3lyp/6-311+g | $C_s$ | -0.982788 | $\mu$:1.29874 $\mu_x$:--0.76338 $\mu_y$:-1.05071 | 300.33 | A:67.14724 B.58.28611 C.58.20919 | 19801 | -2666.844118 |
| b3lyp/ccpvdz | $C_s$ | -0.972217 | $\mu$:1.26996 $\mu_x$:-0.746530 $\mu_y$:-1.027370 | 2069.3 | A:67.29218 B:58. 34914 C:58.22316 | 19774.66 | -2667.20206 |
| pw6b95/ ccpvdz | $C_s$ | -0.967615 | $\mu$:1.32662 $\mu_x$:-0.779835 $\mu_y$:-1.073210 | 816.58 | A:68.15207 B:59.09683 C:58.94779 | 22878.04 | -2671.03709 |
| pbeh1pbe/ ccpvdz | $C_s$ | -0.967646 | $\mu$:1.340260 $\mu_x$:-0.787783 $\mu_y$:-1.084290 | 298.4 | A:67.85503 B:58.83700 C.58.68872 | 21670.9 | -2664.610992 |
| b3pw91/ ccpvdz | $C_s$ | -0.9699822 | $\mu$:1.25694 $\mu_x$:-0.738892 $\mu_y$:-1.01683 | 582.4 | A:67.65895 B:58.66068 C:58.52357 | 19926.1 | -2666.261182 |

The computed quartic centrifugal distortion constans in the Watson asymmetric top representation are very small, a few thousandth of a Hz, except for δK that is about 0.5 Hz. Gaussian 16 yields the effective rotational constants (corrected for quartic centrifugal distortion constants) equal to the equilibrium rotational constants.

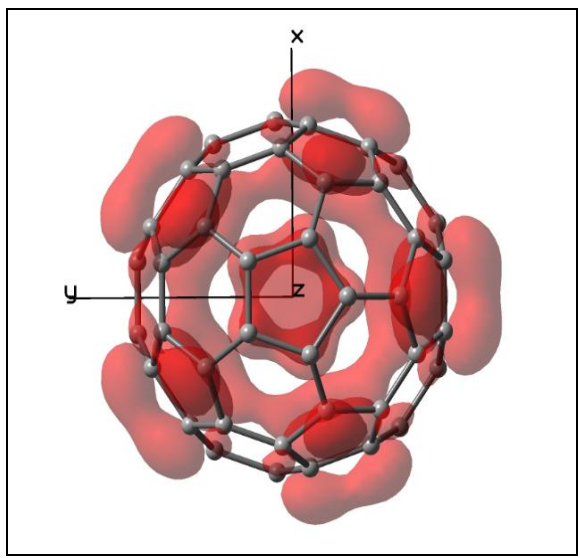

**Figure 1** The HOMO orbital of neutral $C_{70}$ from the b3lyp/6-31+g model has $a_2"$ symmetry

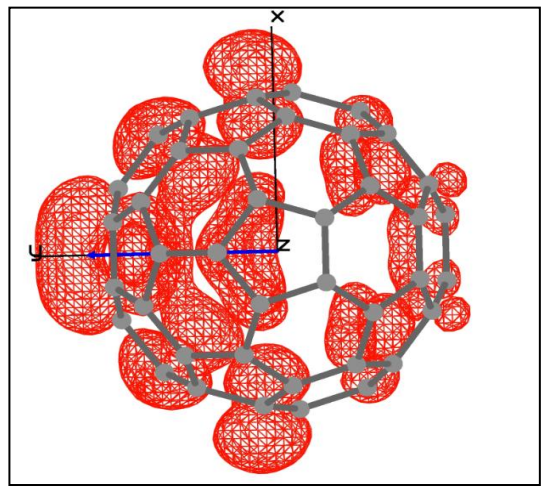

**Figure 2** The positive spin density of the C70+ ion from the pw6b95/ccpvdz model showing the orientation of the blue dipole vector in the y axis

The asymmetry parameter in Table 1. is the Ray parameter defined by the formula:

$$\kappa = (2B-A-C)/(A-C) \qquad (3)$$

in prolate symmetric tops κ = -1. All calculated rotor geometries in Table 1. are seen to be very close to a symmetric top.

A note is appropriate here concerning the orientation of the dipole moment. Table l. shows that the dipole has an x component for all models. The components are reported corresponding to the input geometry used in the calculations, called "spectroscopic orientation" in Gaussian 16, whereas in Figure 2. there is no x component. In Figures 1. and 2. the molecule fixed z axis is perpedicular to the x,y plane. Gaussian 16 computes symmetry more rigorously than Gaussview 6 used for Figures 1. and 2.

## 4. Pseudo Jahn –Teller effect in $C_{70}^+$

In Section 2 a discrepancy was noted with respect to the computed dipole orientation following the Child-Longuet-Higgins symmetry rule. In discussions with Professor Arnout Ceulemans at the Catholic University of Leuven it was found that the very small energy difference between the uppermost two ground state electronic levels $e_1''$ and $a_2''$ (see Figure 3) the HOMO orbital is due to a mixture of these two levels. The molecular orbital of the $e_1''$ level is shown in Figure 4. In Figure 1. HOMO has $a_2''$ symmetry. The symmetry switching between Figures 1. and 4. is due to basis function differences.

This orbital mixing corresponds to a pseudo Jahn-Teller interaction scheme (Bersuker 2006) . The symmetry of the mixed HOMO level is obtained from the direct product of the $D_{5h}$ species $e_1''$ and $a_2''$, which yields $e_1'$. Thus the pseudo Jahn-Teller interaction scheme is in agreement to the induced electric dipole orientation in Table 1,

The sequence of the highest filled and lowest empty orbitals are shown in Figure 3. The yellow orbital is the HOMO, the dotted line separates filled and unfilled orbitals and level energies are given in hartees.

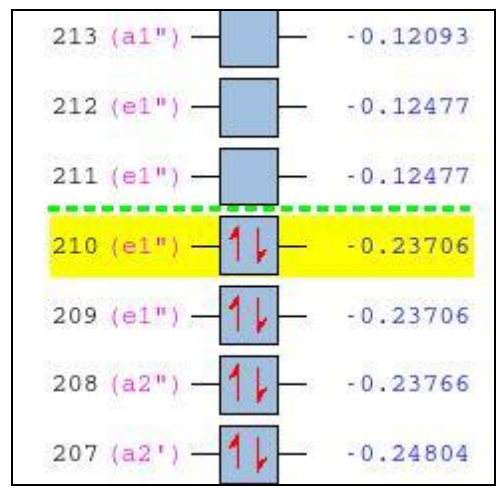

**Figure 3. Orbital energies from the pw6b95/ccpvdz model**

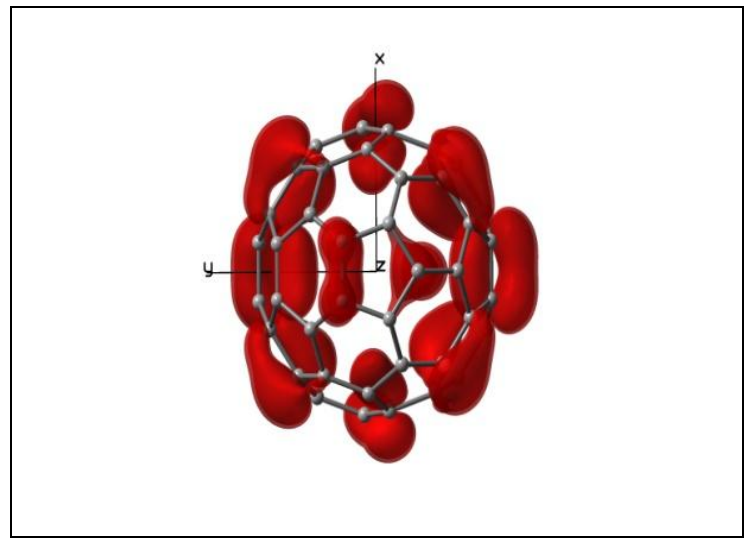

**Figure 4. The HOMO orbital of neutral $C_{70}$ from the pw6b95/ccpvdz model has $e_1''$ symmetry**

## 5. Simulation of the pure rotational radio spectra.

Using the calculated rotational constants and the electric dipole moment components the pure rotational spectrum of $C_{70}^+$ was simulated by the Pgopher software (Western 2017) applying the b3lypccpvdz and pw6b95ccpvdz models. Figures 5.-8. show the calculated spectra for the b3lypccpvdz model at two different temperatures. The spectra for the pw6b95ccpvdz model are very similar to those in Figures 5. - 8.

Table 2. Quartic centrifugal distortion constant (Watson asymmetric top reduction)

| Constants (MHz) | b3lypccpvdz model | pw6b95ccpvdz model |
|---|---|---|
| $\Delta_J$ | 8.2236e-9 | 8.3348e-9 |
| $\Delta_K$ | -3.9094e-9 | -8.8346e-10 |
| $\Delta_{JK}$ | 6.9123e-9 | 3.73590e-9 |
| $\delta_J$ | -1.3577e-10 | -1.0809e-10 |
| $\delta_K$ | -3.9961e-7 | -1.25260e-7 |

The spectra were calculated for temperatures 2.73K and 10K for interstellar and molecular cloud conditions, resp. These spectra contain large number of lines thus Figures 5. and 7. appear as continua. The reason for this is the magnitude of the rotational partition function for high J rotational quantum numbers involved.

The narrow regions in Figures 6, and 8. show regular sequences, containing lines having Einstein A coefficients above the 0.001th portion of the maximum value. The Einstein A coefficient used is the total emission rate from an upper to a lower level on all degenerate M components.

Line width was used to correspond to thermal Doppler width that is 129 Hz at T=3 K and 235 Hz for T=10 K. Natural line width is thus negligible at the these wavelength values. Thus the lines in Figures 6. and 8. have no width.

Although Pgopher provides possibilities for non-Boltzmanian level populations, in absence of any specific radiation transfer model for $C_{70}^+$ the present calculations used Boltzmanian level populations for relative line strengths corresponding to local thermodynamic equilibrium.

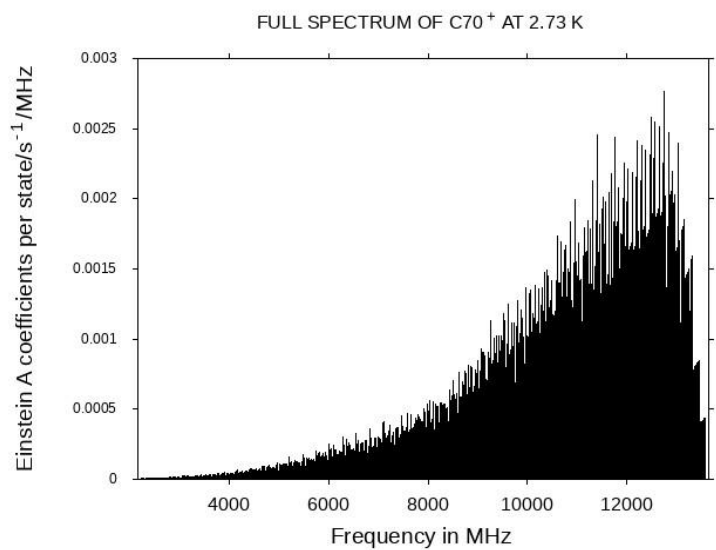

**Figure 5 The rotational spectrum at T=2.73 K with Jmax=100 using the b3lyp/ccpvdz model**

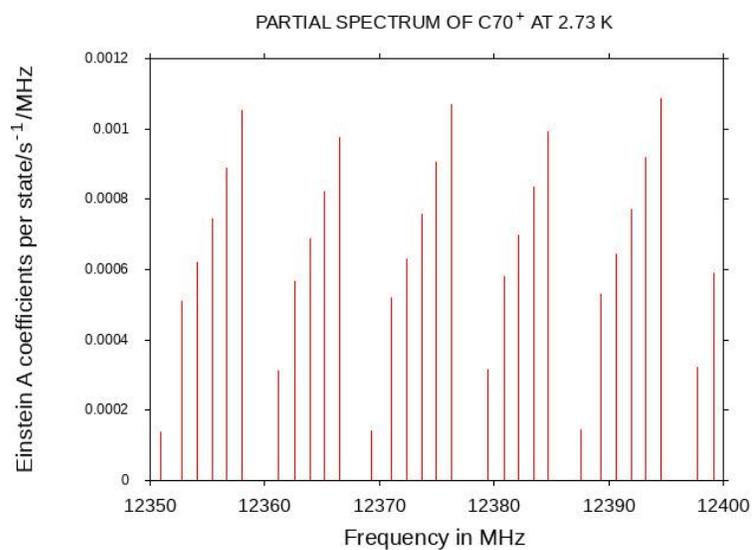

**Figure 6 The partial rotational spectrum at T=2.73 K with Jmax= 100 using the b3lyp/ccpvdz model**

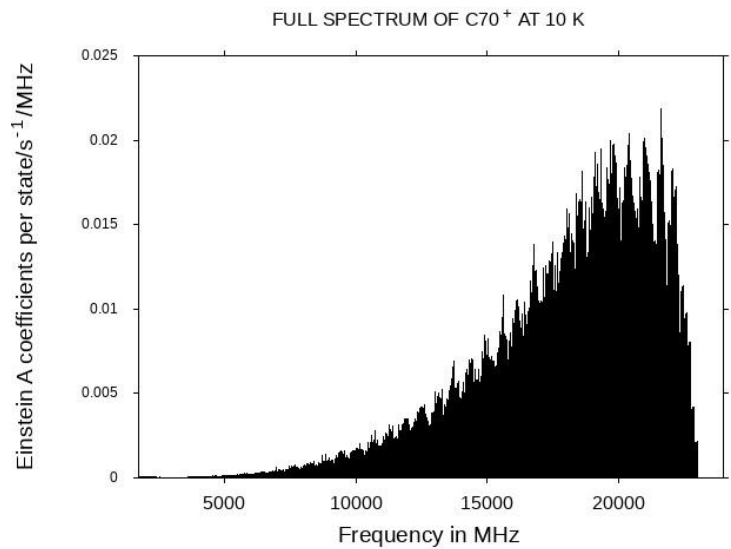

**Figure 7** The rotational spectrum at T=10 K with Jmax= 170 using b3lypccpvdz model

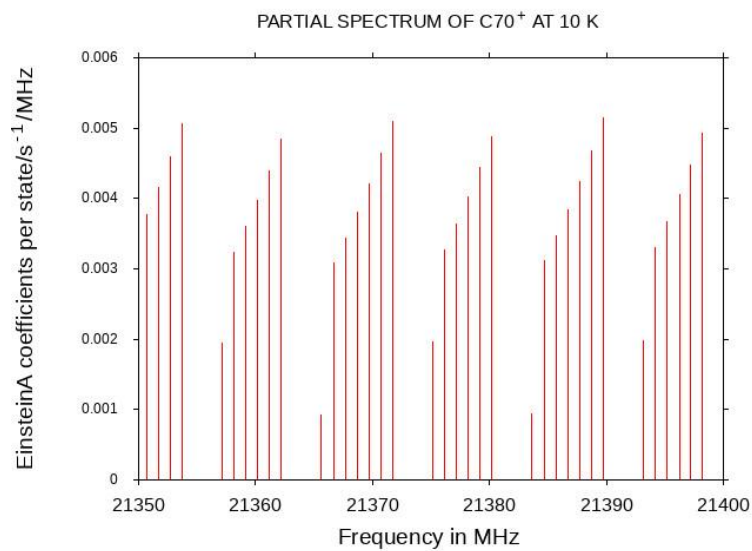

**Figure 8** The partial rotation spectrum at T=10 K with Jmax=170 using b3lypccpvdz model

The resolved lines in Figures 6.. and 8. are all rR-type pure rotational transitions belonging to high J quantum numbers. At T= 2.73 K a maximum of 100 for J corresponds to an almost completely converged rotational partition function, and this is the case also for Jmax=170 in the case of T=10 K.

In the listing of the computed lines in the given frequency range the individual line Einstein coefficients are given, but each line is built from overlapping K components due to the essentially rigid rotor nature of $C_{70}^+$. In addition there are many more lines in the computed spectra that are not shown in the Figures as their intensity falls below the chosen limit, in this case 0.001 times the maximum line intensity.

An important point about the computed spectra is the use of the equilibrium rotational constants. In the laboratory or radio astronomically observed cold rotational spectra the line positions are determined by zero-point vibrational averaged rotational constants. These can be obtained only from anharmonic vibrational calculations, In the case of the heavy $C_{70}$ fullerene with rather stiff C-C bonds ( most C-C bond orders are about 1.5 in the radical cation) mechanical anharmonic averages are probably close to the equilibrium values. The smallness of anharmonicity was stated already using semiempirical calculations for the $C_{60}$ fullerene (Fabian 1996) and was also found in the study of binary combination bands in the hot gas-phase infrared emission spectra of $C_{60}$ and $C_{70}$ (Nemes et al. 1994). Thus the present simulations are probably useful approximation to the actual (so far unavailable) microwave and radio astronomical spectra.

It is also relevant and interesting to note that the wavelength range of the computed spectra overlap with AME (anomalous microwave emission) from diffuse galatic radiation that picks up intensity roughly between 10 and 60 GHz and reaches a maximum at about 22 GHz. It is attributed mainly to emission from spinning ultrasmall interstellar grains (Dickinson . et al 2018).In an earlier work (Iglesias-Groth 2005) AME is alternatively explained as due to electric dipole radiation from rotating hydrogenated spherical fullerenes (single shell fullerenes or multishell bucky onions).

From the present work there appears a possible additional source of AME, the rotational emission from $C_{70}^+$

## .6. Conclusions

The present calculations show for the first time that the $C_{70}+$ radical cation posseses a permanent electrical dipole (about 1.3 Debye) due to static (pseudo) Jahn-Teller effects since ionization reduces molecular symmetry from $D_{5h}$ to $C_s$. Simulation of the pure rotational spectrum based on rotational data and dipole moment from open shell unresticted DFT calculations for two DFT models shows that radio wave transitions cover a range from 2 GHz to 14 GHz at T=2.73 K and from 2 GHz to 24 GHz at T=10 K. Thus there may be possibilities for detecting this fullerene radical ion by laboratory microwave spectroscopy and radio astronomical methods.

Future theoretical Jahn-Teller analyses of the potential hypersurface of $C_{70}^+$ could provide unequivocal proof that global potential minima are reached thus yielding the final answer for molecular geometry. In addion anharmonic vibrational calculations could further refine the present spectral predictions.

## ACKNOWLEDGEMENTS

This work was made possible by the use of the supercomputers of Hungary's Government Information Technology Development Agency (KIFŰ). Special thanks are due to Dr. Attila Fekete for his guidence. The author is indebted to Dr. Douglas J. Fox, Director at the Gaussian Corporation. Wallingford, Connecticut, US. for his extensive help with Gaussian 16.

In the course of this work discussions with several colleagues have been of great value. In particular I was helped by the suggestions of the following scentists: Professor Jan Cami, University of Western Ontario, Canada, Dr. D.A. Garcia – Hernandez, Istituto de Astrofisica de Canarias, Spain, Professor Arnout Ceulemans, University of Leuven, Leuven. Belgium, Professor Dmitry Strelnikov, Institute of Physical Chemistry, Karlsruhe, Germany, Dr. Koichi M.T. Yamada, NAIR, Tsukuba, Japan, and Dr. Stephan Irle, Oak Ridge, National Laboratory, USA.

# Supplementary material

Computed pure rotational spectra ascii files using the b3lyp/ccpvdz and pw6b95/ccpvdz models for T=2.73 K and T=10K temperatures are available from the author

Yamada K.M.T., Ross S.C., Fumiyuki I., 2017, Mol. Astrophys. 6, 9